\documentclass{article}

\PassOptionsToPackage{numbers, compress}{natbib}

\usepackage[preprint]{neurips_2023}

\usepackage[utf8]{inputenc} 
\usepackage[T1]{fontenc}    
\usepackage{hyperref}       
\usepackage{url}            
\usepackage{booktabs}       
\usepackage{amsfonts}       
\usepackage{nicefrac}       
\usepackage{microtype}      
\usepackage{xcolor}         

\usepackage{multirow}
\usepackage{array,makecell,booktabs}
\newcolumntype{M}[1]{>{\centering\arraybackslash}m{#1}}
\usepackage{wrapfig}

\usepackage{graphicx}
\usepackage{float}
\usepackage{overpic}
\usepackage{wrapfig}
\usepackage{caption}
\usepackage{subfigure}

\usepackage{amsmath}

\usepackage{bbding}

\usepackage{appendix}
\usepackage{enumitem}

\title{Mega-TTS: Zero-Shot Text-to-Speech at Scale \\ with Intrinsic Inductive Bias}

\author{
Ziyue Jiang~\thanks{Equal contribution.}~~\thanks{Interns at ByteDance.}~~$^{\spadesuit\heartsuit}$
~~Yi Ren~\footnotemark[1]~~$^{\heartsuit}$
~~Zhenhui Ye~\footnotemark[1]~~\footnotemark[2]~~$^{\spadesuit\heartsuit}$ 
~~Jinglin Liu~$^{\heartsuit}$
~~Chen Zhang~\footnotemark[2]~~$^{\heartsuit\spadesuit}$ \\
\textbf{
~~Qian Yang~$^\spadesuit$
~~Shengpeng Ji~$^\spadesuit$ 
~~Rongjie Huang~$^\spadesuit$ 
~~Chunfeng Wang~$^{\heartsuit}$ 
} \\
\textbf{
~~Xiang Yin~$^{\heartsuit}$
~~Zejun Ma~$^{\heartsuit}$ 
~~Zhou Zhao~\thanks{Corresponding author.}~~$^\spadesuit$
} \\
$^\spadesuit$Zhejiang University \& $^{\heartsuit}$ByteDance \\
\texttt{ziyuejiang@zju.edu.cn, ren.yi@bytedance.com, zhaozhou@zju.edu.cn} \\
}

\begin{document}

\maketitle

\begin{abstract}

Scaling text-to-speech to a large and wild dataset has been proven to be highly effective in achieving timbre and speech style generalization, particularly in zero-shot TTS. However, previous works usually encode speech into latent using audio codec and use autoregressive language models or diffusion models to generate it, which ignores the intrinsic nature of speech and may lead to inferior or uncontrollable results. We argue that speech can be decomposed into several attributes (e.g., content, timbre, prosody, and phase) and each of them should be modeled using a module with appropriate inductive biases. From this perspective, we carefully design a novel and large zero-shot TTS system called Mega-TTS, which is trained with large-scale wild data and models different attributes in different ways: 1) Instead of using latent encoded by audio codec as the intermediate feature, we still choose spectrogram as it separates the phase and other attributes very well. Phase can be appropriately constructed by the GAN-based vocoder and does not need to be modeled by the language model. 2) We model the timbre using global vectors since timbre is a global attribute that changes slowly over time. 3) We further use a VQGAN-based acoustic model to generate the spectrogram and a latent code language model to fit the distribution of prosody, since prosody changes quickly over time in a sentence, and language models can capture both local and long-range dependencies. We scale Mega-TTS to multi-domain datasets with 20K hours of speech and evaluate its performance on unseen speakers. Experimental results demonstrate that Mega-TTS surpasses state-of-the-art TTS systems on zero-shot TTS, speech editing, and cross-lingual TTS tasks, with superior naturalness, robustness, and speaker similarity due to the proper inductive bias of each module. Audio samples are available at \url{https://mega-tts.github.io/demo-page}.



\end{abstract}

\section{Introduction}
Text-to-speech (TTS) synthesis~\citep{tan2021survey,arik2017deep,shen2018natural,li2019neural,ren2019fastspeech,ren2020fastspeech,kim2020glow,zhang2021denoispeech,popov2021grad,kim2022guided} aims to generate human-like speech from text and has gained significant attention in the field of machine learning. Traditional TTS systems~\citep{chen2018sample,chen2021adaspeech,wang2018style,casanova2022yourtts,huang2022generspeech} are usually trained on limited datasets, which impairs their models' ability to produce diverse and generalizable results. In contrast, large-scale TTS systems~\citep{wang2023neural,zhang2023speak,kharitonov2023speak} are trained on tens of thousands of hours of speech data, which significantly improves their zero-shot capability~\citep{wang2023neural,zhang2023speak}. Current large-scale TTS systems typically encode the speech waveform into latent with neural codec models~\citep{defossez2022high} as the intermediate representation and model it with autoregressive language models (LM)~\citep{wang2023neural} or diffusion models~\citep{shen2023naturalspeech}.

\begin{table}[t]
\caption{The illustration of intrinsic properties for different components in human speech.}
\label{table_1}
\centering
\begin{tabular}{@{}l|c|c|c@{}}
\toprule
Modality & Components & Intrinsic Properties & Suitable for LM    \\                     
\midrule
\multirow{6.8}{*}{Human speech} & Phase & Highly dynamic, irrelevant to semantics & \XSolidBrush \\
\cmidrule(l){2-4} & Timbre & Global and stable & \XSolidBrush\\
\cmidrule(l){2-4} & \multirow{3}{*}{Prosody} & Long-term dependencies & \multirow{3}{*}{\Checkmark} \\
& & Rapid changes &  \\
& & Weak relation with text &  \\
\cmidrule(l){2-4} & Content & Monotonic alignment & \XSolidBrush\\
\bottomrule
\end{tabular}
\end{table}

As presented in Table~\ref{table_1}, human speech can be decoupled into several attributes: content, timbre, prosody, phase, etc. However, current large-scale TTS systems directly use neural audio codec models to encode the entire speech into latent and ignore the following intrinsic nature of speech: 1) phase is highly dynamic and irrelevant to semantics, which means people are much less sensitive to perceive phase than to prosody and timbre, especially for monaural audio. Therefore, only one reasonable phase is needed for waveform reconstruction, and it is not necessary to model all possible phases. Modeling phase with LM or diffusion model can waste a lot of model parameters since they model the full distribution of phase\footnote{That is why GAN-based vocoders~\cite{kong2020hifi} are popular.}. 2) Timbre should remain stable within the sentence as a global vector. Modeling timbre with time-varying latent is costly\footnote{Our method retains a small portion of time-varying timbre information in the latent code, while the majority is represented as the global vector.}. 3) Prosody typically has both local and long-term dependencies and changes rapidly over time with a weak correlation to text, which makes conditional phoneme-level LLMs inherently ideal for generating prosody sequences. 4) Content has monotonic alignment with speech while the autoregressive language model cannot guarantee that, which can lead to repeating or missing word issues~\citep{wang2017tacotron,wang2023neural,zhang2023speak}.

To make use of the large and wild training datasets while matching the inductive bias of the model and the intrinsic nature of speech, we propose a zero-shot text-to-speech model called Mega-TTS. Specifically, 1) considering the limitations of neural audio codec models, we select mel-spectrogram as the intermediate representation to separate the phase and other attributes. We adopt a GAN-based vocoder to reconstruct the phase information to improve our model's efficiency. 2) To model timbre information, we employ global vectors since timbre is a global attribute that changes slowly over time. We extract the global information from a different speech of the same speaker with the global speaker encoder to decompose the timbre and content information. 3) To capture prosody information in a sentence, we adopt a VQGAN-based acoustic model to generate the mel-spectrogram and a latent code language model called P-LLM to fit the distribution of prosody. The P-LLM is capable of capturing both local and long-range dependencies for prosody modeling.

To evaluate the zero-shot performance of Mega-TTS, we perform experiments on VCTK~\citep{veaux2016superseded}, AISHELL-3~\citep{shi2020aishell} and LibriSpeech test-clean~\citep{panayotov2015librispeech} datasets. All of the test speakers are unseen in the training corpus. Our Mega-TTS surpasses the state-of-the-art zero-shot TTS systems~\citep{casanova2022yourtts,wang2023neural} in terms of speaker similarity, speech naturalness, and generation robustness, which demonstrates the superiority of introducing appropriate inductive biases. Moreover, Mega-TTS outperforms state-of-the-art models on speech editing~\citep{tan2021editspeech,bai20223} and cross-lingual TTS~\citep{zhang2023speak} tasks. The main contributions of this work are summarized as follows:

\begin{itemize}[leftmargin=*]

\item We propose Mega-TTS, a zero-shot text-to-speech system that considers intrinsic inductive biases. Instead of using latent encoded by audio codec as the intermediate representation~\citep{zeghidour2021soundstream,defossez2022high,wang2023neural}, we decompose mel-spectrogram into content, timbre, prosody, and phase attributes and model each of them according to their intrinsic properties.

\item We train Mega-TTS on a multi-domain and multi-lingual dataset that contains 20k hours of speech data. It is worth noting that existing large-scale TTS systems~\citep{wang2023neural,shen2023naturalspeech} are typically trained with speech corpora from audiobooks, while our system is trained on multi-domain speech corpora.

\item We evaluate Mega-TTS on 3 down-stream speech generation tasks (i.e., zero-shot TTS, speech editing, and cross-lingual TTS), demonstrating that Mega-TTS can be applied to various speech generation tasks. We also propose a novel sampling strategy for speech editing via the discrete prosody tokens extracted by Mega-TTS. 
\end{itemize}

\section{Background}
In this section, we briefly overview the background of this work, including zero-shot text-to-speech (TTS) and generative models for speech synthesis.

\paragraph{Zero-shot text-to-speech.}
Text-to-speech models usually generate mel-spectrogram from text~\citep{wang2017tacotron,arik2017deep,li2019neural,ren2019fastspeech,kim2020glow,ren2021portaspeech,liu2022diffsinger,huang2022prodiff} and then synthesize speech waveform from the generated mel-spectrogram using a separately pre-trained vocoder~\citep{oord2016wavenet,kong2020hifi,yamamoto2020parallel,huang2022fastdiff}, or directly generate waveform from text in an end-to-end manner~\citep{ren2020fastspeech,DBLP:conf/iclr/DonahueDBES21,kim2021conditional,liu2022delightfultts}.
For decades, the increasing demand for personalized speech generation in various applications has posed challenges for TTS models~\citep{tan2021survey}, especially in zero-shot multi-speaker scenarios regarding domain shifts. Previous approaches can be categorized into speaker adaptation~\citep{chen2018sample,chen2021adaspeech,wang2018style,huang2022meta} and speaker encoding~\citep{jia2018transfer,arik2018neural,kang2022any,wu2022adaspeech} methods. Traditional works are typically trained on small datasets~\citep{chen2021adaspeech,huang2022meta,huang2022generspeech,casanova2022yourtts}, while some recent works~\citep{borsos2022audiolm,wang2023neural,kharitonov2023speak,zhang2023speak} are trained on large-scale datasets and demonstrate the effectiveness in zero-shot scenarios. These systems utilize the neural audio codec models~\citep{zeghidour2021soundstream,defossez2022high} to convert audio waveform into latent and consider it as the intermediate representation for speech generation. Among them, SPEAR-TTS~\citep{kharitonov2023speak} splits the TTS task into two sequence-to-sequence tasks, which enables the training using abundant audio-only data. NaturalSpeech 2~\citep{shen2023naturalspeech} uses a text-conditioned diffusion model to generate the latent vectors of the neural audio codec model. VALL-E~\citep{wang2023neural,zhang2023speak} proposes the first neural codec language model for text-to-speech, exhibiting strong in-context learning abilities to overcome challenges in zero-shot speech generation. However, these methods ignore the intrinsic property of speech and may lead to inferior or uncontrollable results (e.g., word skipping, repeating, and collapse~\citep{wang2023neural,zhang2023speak}). Considering the nature of different speech attributes, the autoregressive language model is ideally suitable for prosody modeling. ProsoSpeech~\citep{ren2022prosospeech} has proposed to improve the prosody modeling for TTS with latent prosody vectors predicted by a language model. Nevertheless, it lacks the in-context learning capacity, which restricts its application scenarios.

\paragraph{Generative models for speech synthesis.}
Generative models, like language models~\citep{borsos2022audiolm,kreuk2022audiogen}, VAE~\citep{lee2021bidirectional,ren2021portaspeech}, GAN~\citep{kong2020hifi,kim2021conditional}, Normalizing flow~\citep{miao2020flow,kim2020glow}, and diffusion model~\citep{kong2020diffwave,jeong2021diff,popov2021grad,huang2022prodiff}, have been applied to speech or audio synthesis for years. Previous works of autoregressive generative model mainly aim at waveform generation~\citep{oord2016wavenet,goel2022s} and continuous acoustic feature generation~\citep{wang2017tacotron,shen2018natural}. Recently, speech generation systems like AudioLM~\citep{borsos2022audiolm} and VALL-E~\citep{wang2023neural} propose to utilize neural audio codec models~\citep{zeghidour2021soundstream,defossez2022high} to convert audio waveform into discrete codes as the intermediate representation and design LLMs to generate these codes to achieve speech synthesis. Although good reconstruction quality can be achieved by neural audio codec models, they ignore the intrinsic nature of speech~\citep{defossez2022high} and may not be suitable to serve as the generator of intermediate representation for speech generation. The encoded latent contains the phase, content, and timbre attributes and language models are not suitable for predicting these due to the error propagation problem.

\section{Method}
\begin{figure*}[!t]
    \centering
    \includegraphics[width=.98\linewidth]{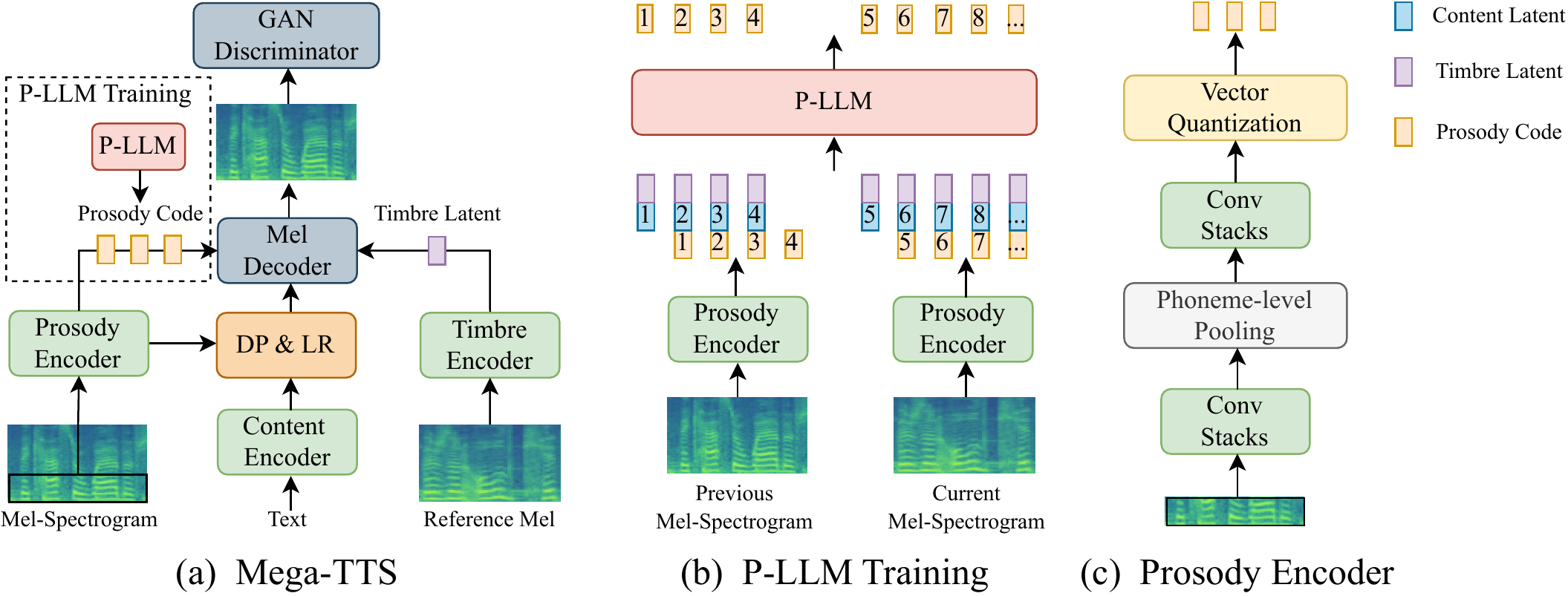}
\caption{The overall architecture for Mega-TTS. In subfigure (a), P-LLM denotes the prosody large language model; DP \& LR denote the duration predictor and length regulator proposed in FastSpeech~\citep{ren2019fastspeech}. In subfigure (b), P-LLM autoregressively predicts the discrete prosody codes.}
\label{main_architecture}
\end{figure*}

To introduce proper inductive biases into large-scale TTS systems, we propose Mega-TTS, a zero-shot TTS system for natural and robust speech generation in various scenarios (i.e., zero-shot prompt-based TTS, speech editing, and cross-lingual TTS). As shown in Figure~\ref{main_architecture}, Mega-TTS consists of a VQGAN-based~\citep{esser2021taming} TTS model and a prosody large language model (P-LLM). We carefully model different speech attributes in different ways. First, we choose the mel-spectrogram as the intermediate representation as it separates the phase from other attributes very well. Secondly, we extract the global vector from the random previous sentence of the same speaker with the global timbre encoder to disentangle the timbre and content information. Finally, we further use a VQGAN-based acoustic model to generate the mel-spectrogram and propose a latent code language model called P-LLM to fit the distribution of prosody, since language models are capable of capturing both local and long-range dependency. During inference, we propose to use the content from the given text sequence, the timbre extracted from the prompt speech, and the prosody predicted by our P-LLM to generate the target speech, which is a novel TTS decoding mechanism called \textit{prosody-oriented speech decoding}. Finally, to demonstrate that our model can be applied to various scenarios, we design inference strategies for downstream tasks. We describe these designs and the training and inference procedures in detail in the following subsections.

\subsection{Disentangling speech into different components}
\label{sec_3_1}
To introduce appropriate inductive biases into different speech attributes, we need to separately express these attributes and carefully design different architectures for them. The overall model architecture of Mega-TTS is shown in Figure~\ref{main_architecture}. We use three types of encoders to separately encode content, prosody, and timbre representations. Then we adopt a GAN-based mel-spectrogram decoder to generate mel-spectrograms with these representations. We describe the disentangling strategy and detailed design of the proposed encoders as follows.

\paragraph{Disentangling strategy.} We disentangle the mel-spectrogram into content, prosody, and timbre representations with the reconstruction loss of the autoencoder and a carefully designed bottleneck~\citep{qian2019autovc}: 1) we feed the mel-spectrogram into the prosody encoder, and we also introduce carefully-tuned dimension reduction and phoneme-level downsampling to the prosody encoder to constrain the information flow;
2) the content encoder encodes the phoneme sequence into the content representation;
3) we feed the reference mel-spectrogram sampled from a different speech of the same speaker to disentangle the timbre and content information and temporally average the output of the timbre encoder to get a one-dimensional global timbre vector. The correctly-designed bottleneck will learn to remove the content information and the global timbre information from the output of the prosody encoder, which ensures the performance of disentanglement. Due to the limited page space, we put more details about the hyperparameter selection for the information bottleneck in Appendix~\ref{app:hyperparameter_selection_for_bn}.

\paragraph{Architecture design of encoders.} 1) \textit{The prosody encoder} consists of two convolution stacks, a phoneme-level pooling layer, and a vector quantization (VQ) bottleneck. The first convolution stacks compress mel-spectrograms into phoneme-level hidden states according to the phoneme boundary and the second stacks capture phoneme-level correlations. The vector quantization layer~\citep{van2017neural} then utilizes these hidden states to obtain phoneme-level prosody codes $\mathbf{u} = \{u_{1},u_{2},...,u_{T}\}$ and hidden states $H_{prosody}$. To ease the difficulty of disentanglement, only the low-frequency band of the mel-spectrogram (the first 20 bins in each mel-spectrogram frame) is used as input, as it contains almost complete prosody and significantly less timbre/content information compared to the full band~\citep{ren2022prosospeech}; 2)  \textit{The content encoder} is composed of several feed-forward Transformer layers. To achieve the monotonic alignment between the speech content and generated speech, we adopt the duration predictor and length regulator following common practice in non-autoregressive TTS systems~\citep{ren2019fastspeech,shen2023naturalspeech}. Differently, we feed the prosody information extracted by the prosody encoder to the duration predictor in order to ease the one-to-many mapping problem~\citep{ren2019fastspeech,ren2020fastspeech}; 3) \textit{The timbre encoder} is designed to extract a global vector $H_{timbre}$ that contains the speaker identity of the given speech. The timbre encoder consists of several stacks of convolution layers. To ensure the stability of timbre information across the time axis, we temporally average the output of the timbre encoder to get a one-dimensional timbre vector $H_{timbre}$. 

To keep good perceptual quality, we introduce a GAN-based mel-spectrogram decoder. We adopt the multi-length discriminator~\citep{chen2020hifisinger,ye2022syntaspeech} based on random windows of different lengths as the discriminator. Overall, the first-stage training loss $\mathcal{L}$ of Mega-TTS can be formulated as:
\begin{equation}
\mathcal{L}_{\mathrm{VQ}} = \|y_{t}-\hat{y}_{t}\|^2+\left\|\operatorname{sg}[E(y_{t})]-z_{\mathbf{q}}\right\|_2^2+\left\|\operatorname{sg}\left[z_{\mathbf{q}}\right]-E(y_{t})\right\|_2^2\ ,
\end{equation}
\begin{equation}
\mathcal{L}=\mathbb{E}\left[ \mathcal{L}_{\mathrm{VQ}} + \mathcal{L}_{\mathrm{Adv}}\right]\ ,
\end{equation}
where $y_{t}$ is the target speech and $\hat{y}_{t}$ is the generated speech. $\mathcal{L}_{\mathrm{rec}}=\|y_{t}-\hat{y}_{t}\|^2$ is the reconstruction loss, $\operatorname{sg}[\cdot]$ denotes the stop-gradient operation, and $z_{\mathbf{q}}$ is the temporal collection of codebook entries. $\mathcal{L}_{\mathrm{VQ}}$ is the VQVAE loss function~\citep{van2017neural,esser2021taming} and $\mathcal{L}_{\mathrm{Adv}}$ is the LSGAN-styled adversarial loss~\citep{mao2017least} whose objective is to minimize the distribution distance between the predicted mel-spectrograms and the ground truth mel-spectrograms.





\subsection{P-LLM}
\label{P-LLM_Training}
The P-LLM is a latent code language model that captures local and long-range dependency for prosody modeling. We describe the prosody-oriented speech decoding mechanism and details of the P-LLM as follows.

\paragraph{Prosody-oriented speech decoding.} Denote $(\mathbf{y_{p}}, \mathbf{x_{p}})$ and $(\mathbf{y_{t}}, \mathbf{x_{t}})$ as the prompt and target speech-transcription pairs. Our goal is to synthesize the high-quality target speech $\mathbf{y_{t}}$ given an unseen speech prompt $\mathbf{y_{p}}$. During inference, the timbre of the target speech $\tilde{H}_{timbre}$ is expected to be the same as that of the prompt speech. Therefore, to generate the target speech $\mathbf{y_{t}}$, we only need the prosody information $\tilde{\mathbf{u}}$ of the target speech. Therefore, the prosody-oriented speech decoding procedure can be formulated as follows:
\begin{equation}
\begin{aligned}
&\textbf{Encode}: \mathbf{u}=E_{prosody}(\mathbf{y_{p}}),\ 
H_{content} = E_{content}(\mathbf{x_{p}}),\ 
\tilde{H}_{timbre} = E_{timbre}(\mathbf{y_{p}}),\ \\
&\quad\quad\quad\ \ \ \tilde{H}_{content} = E_{content}(\mathbf{x_{t}})\ ,\\
&\textbf{Prosody prediction}: \tilde{\mathbf{u}} = f(\tilde{\mathbf{u}}|\mathbf{u}, H_{content}, \tilde{H}_{timbre}, \tilde{H}_{content}; \theta)\ ,\\
&\textbf{Decode}: \hat{y}_{t} = D(\tilde{\mathbf{u}},\tilde{H}_{timbre},\tilde{H}_{content})\ ,
\end{aligned}
\end{equation}
where $E_{prosody}$, $E_{timbre}$, $E_{content}$, and $D$ denote the prosody encoder, timbre encoder, content encoder, and mel decoder. $\mathbf{u}$ is the prosody tokens of the prompt speech, $\tilde{\mathbf{u}}$ is the predicted prosody tokens of the target speech, $f$ is the prosody prediction function, and $\theta$ is the parameter of the P-LLM. $\hat{y}_{t}$ is the generated speech. 

\paragraph{Generating prosody codes.} The proposed prosody-oriented speech decoding mechanism requires the predicted prosody codes $\tilde{\mathbf{u}}$ of the target speech. Leveraging the powerful in-context learning capability of LLMs, we design the P-LLM module to predict $\tilde{\mathbf{u}}$. The P-LLM is a decoder-only transformer-based architecture~\citep{brown2020language} for prosody modeling, which uses prosody codes $\mathbf{u}$ from $\mathbf{y_{p}}$ as the prompt and $H_{content}$, $\tilde{H}_{content}$, and $\tilde{H}_{timbre}$ as the condition. The autoregressive prosody prediction process of P-LLM can be formulated as:
\begin{equation}
   p\left(\tilde{\mathbf{u}} \mid \mathbf{u}, H_{content}, \tilde{H}_{timbre}, \tilde{H}_{content} ; \theta \right)=\prod_{t=0}^T p\left(\tilde{u}_{t} \mid \tilde{u}_{<t}, \mathbf{u}, H_{content}, \tilde{H}_{timbre}, \tilde{H}_{content} ; \theta \right),
   \label{eq_4}
\end{equation}
where $\theta$ is the parameter of our P-LLM. Since the discrete prosody sequence $\mathbf{u}$ is phoneme-level, we directly concatenate it with $H_{content}$, $\tilde{H}_{content}$, and $\tilde{H}_{timbre}$ as the input. The P-LLM is trained in a teacher-forcing mode in the training stage via the cross-entropy loss.

\begin{figure*}[t]
    \centering
    \includegraphics[width=.90\linewidth]{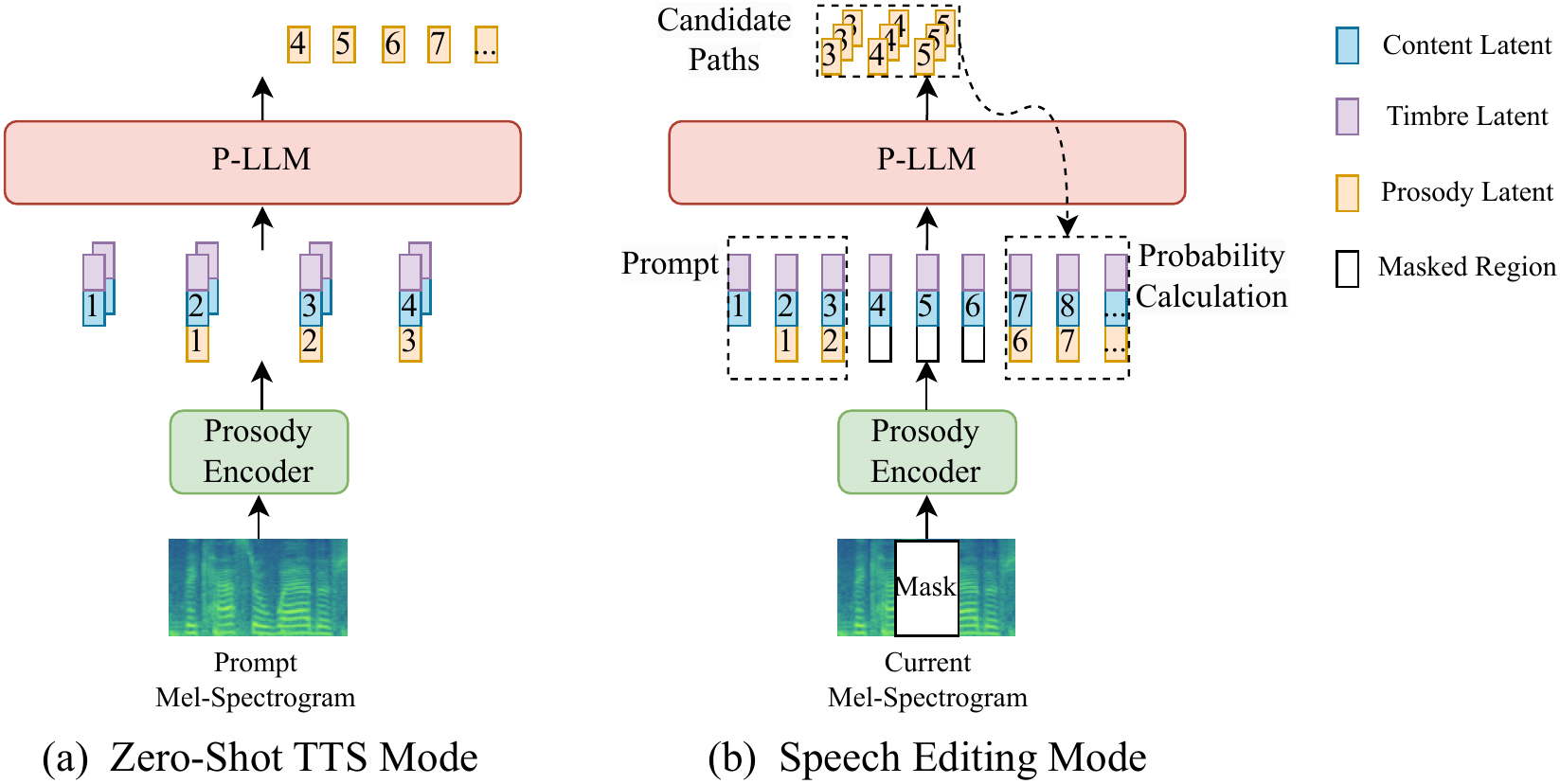}
\caption{The inference modes for Mega-TTS. In subfigure (a), P-LLM takes information from the prompt speech to generate prosody codes for the target speech; In subfigure (b), P-LLM utilizes the contextual information from the masked speech to perform speech editing.}
\label{infernce_mode}
\end{figure*}

\subsection{Speech prompting for inference}
To facilitate in-context learning for various speech generation tasks, we design different speech prompting mechanisms to encourage Mega-TTS to follow the information in the speech prompt.

\paragraph{Inference for TTS.} For zero-shot TTS, P-LLM uses $\mathbf{u}, H_{content}, \tilde{H}_{timbre}, \tilde{H}_{content}$ to generate the target prosody codes $\tilde{\mathbf{u}}$ for the target speech according to Equation~\ref{eq_4}. We use the top-k
random sampling scheme~\citep{fan2018hierarchical} to sample the results since we observe that the sampling-based method could increase the diversity of the generated speech. Then, we concatenate the content $\tilde{H}_{content}$, timbre $\tilde{H}_{timbre}$, and prosody $\tilde{\mathbf{u}}$ information to generate the target speech $y_{t}$ using the mel decoder. Leveraging the proper inductive biases and powerful in-context learning capability of our P-LLM, the generated speech can retain not only similar timbre but also the rhythmic habits of the prompt speech. For cross-lingual TTS, $\mathbf{u}, H_{content}, \tilde{H}_{timbre}, \tilde{H}_{content}$ are extracted from the prompt speech in a foreign language, and the subsequent procedure keeps the same as that of zero-shot TTS.

\paragraph{Inference for speech editing.} In speech editing, the predicted prosody codes should achieve smooth transitions at both the left and right boundaries of the masked region. Previous works like EditSpeech~\citep{tan2021editspeech} propose to perform left and right autoregressive inferences separately and concat the mel-spectrogram at the least L2-norm difference fusion point. However, the L2-norm difference of the mel-spectrogram is far from human perception, leading to poor audio naturalness. Since the prosody representations in Mega-TTS is discrete, we can solve the transition problem by operating on discrete prosody representations. First, we regard the area on the left side of the mask as a prompt to generate $N$ candidate paths with top-k random sampling strategy. Secondly, the $N$ generated paths are used as new prompts to generate the probability matrix of the area on the right side of the mask and the ground-truth prosody codes are used to obtain the probabilities of each decoding step from the probability matrix. In the third stage, we sum up the log probabilities of each decoding step for the candidate paths. Finally, we choose the path that achieves the maximum probability in the second step as the predicted result. The decoding strategy for speech editing can be formulated as follows: \begin{equation}
\begin{aligned}
    \mathop{\text{Max}}_{i\in[1,N]} \text{Likelihood}=
    &\mathop{\text{Max}}_{i\in[1,N]} \prod_{t=L}^R p\left(u_{t}^{i} \mid u_{<t}^{i}, H_{content}, \tilde{H}_{timbre}, \tilde{H}_{content} ; \theta \right) \\
    &\cdot \prod_{t=R}^{T} p\left(u_{t}^{gt} \mid u_{<t}^{i}, H_{content}, \tilde{H}_{timbre}, \tilde{H}_{content} ; \theta \right),
\label{eq_5}
\end{aligned}
\end{equation}
where $L$ and $R$ are the left and right boundaries of the mask. $T$ is the length of the mel-spectrogram. $u^{i}$ is the prosody code in the i-th candidate path. $u_{t}^{gt}$ is the ground-truth prosody codes. Since our decoding strategy considers the prosody information of the boundaries on both sides, the edited region can achieve smooth transitions.
\label{P-LLM_Infernce}

\section{Experiments}
\label{sec_4}
In this section, we present the evaluation results of Mega-TTS and the comparison with baselines in terms of the objective and subjective metrics.
\subsection{Experimental setup}
\label{Experimental_Setup}
\paragraph{Training datasets.} We use GigaSpeech~\citep{chen2021gigaspeech} and WenetSpeech~\citep{zhang2022wenetspeech} as the training corpora, which contains 20k hours of multi-domain speeches in English and Chinese in total. Since the speech in GigaSpeech and WenetSpeech does not have speaker identities and multiple speakers may appear in a speech clip, we process the datasets with an open-source automatic speaker diarization model\footnote{\url{https://huggingface.co/pyannote/speaker-diarization}}~\citep{Bredin2020,Bredin2021}. We also extract the phoneme-level alignments with the external alignment tool\footnote{\url{https://github.com/MontrealCorpusTools/Montreal-Forced-Aligner}}. More information can be found in Appendix~\ref{app:diarization_model}.


\paragraph{Evaluation datasets.} We employ two datasets for evaluation: 1) VCTK dataset~\citep{veaux2016superseded}, an English dataset that contains 108 speakers; 2) LibriSpeech~\citep{panayotov2015librispeech} test-clean, an English dataset that contains 40 speakers. For each of these datasets, we randomly sample 10 utterances for each of the 40 speakers, resulting in a subset of 400 utterances for evaluation; Specifically, to synthesize each sample, we randomly select a different utterance of the same speaker to form the speech prompt. Note that all speakers in the evaluation datasets are unseen during training.

\paragraph{Model configuration.} 
Our Mega-TTS consists of three encoders, a prosody large language model, a mel decoder, and a discriminator. The prosody encoder, timbre encoder, and mel generator consist of 5 convolutional blocks with 320 hidden size, 5 convolution 1D kernel size. The content encoder is a 4-layer Transformer~\citep{vaswani2017attention} with 2 attention heads, 320 embedding dimensions, 1280 1D convolution filter size, and 5 convolution 1D kernel size. The duration predictor is a 3-layer 1D convolution with ReLU activation and layer normalization, which have 320 hidden size. The discriminator follows the architecture proposed in SyntaSpeech~\citep{ye2022syntaspeech}. The P-LLM model is a decoder-only architecture that contains 8 Transformer layers with 8 attention heads, 512 embedding dimensions, 2048 1D convolution filter size, and 5 convolution 1D kernel size. The overall number of model parameters is 222.5M. We add more detailed model configurations in Appendix~\ref{megatts_config}.

\paragraph{Training and inference.} 
In the training stage, we train Mega-TTS on 8 NVIDIA A100 GPUs, with a batch size of 30 sentences on each GPU. We use the Adam optimizer with $\beta_1 = 0.9$, $\beta_2 = 0.98$, $\epsilon = 10^{-9}$ and follow the same learning rate schedule in~\cite{vaswani2017attention}. It takes 320k steps for the VQ-GAN TTS model's training and 100K steps for the P-LLM's training until convergence. The predicted mel-spectrograms are transformed into audio samples using pre-trained HiFi-GAN V1\footnote{\url{https://github.com/jik876/hifi-gan}}~\cite{kong2020hifi}. In the inference stage, we use the top-5 random sampling scheme ~\citep{fan2018hierarchical} to sample diverse results.

\paragraph{Objective metrics.} We evaluate the pitch distance and speaker similarity for zero-shot TTS. In terms of the pitch distance, we compute the average dynamic time warping (DTW)~\cite{muller2007dynamic} distances between the pitch contours of ground-truth speech and synthesized speech. And for the cosine speaker similarity, we use the WavLM model~\citep{chen2022wavlm} finetuned for speaker verification\footnote{\url{https://huggingface.co/microsoft/wavlm-base-plus-sv}} to compute the cosine speaker similarity score between the ground-truth speech and synthesized speech. The similarity score is in the range of $\left[-1,1\right]$, where a larger value indicates a higher similarity of input samples. In addition, we also evaluate the word error rate (WER) for cross-lingual TTS. We use the ASR system from the released HuBERT-Large model~\citep{hsu2021hubert} to transcribe the generated speech into text. Then, the WER between the transcribed text and the original target text is measured. We use all samples in the test set for the objective evaluation. We put more information in Appendix~\ref{app:details_speaker_similarity_model} and Appendix~\ref{app:details_asr_model}. 

\paragraph{Subjective metrics.} We conduct the MOS (mean opinion score) and CMOS (comparative mean opinion score) evaluation on the test set to measure the audio naturalness via Amazon Mechanical Turk. We keep the text content and prompt speech consistent among different models to exclude other interference factors. We randomly choose 50 samples from the test set of each dataset for the subjective evaluation and each audio is listened to by at least 20 testers. We analyze the MOS in three aspects: MOS-Q (Quality: clarity, high-frequency, and original timbre reconstruction), MOS-P (Prosody: naturalness of pitch, energy, and duration), and MOS-S (Speaker similarity). We also analyze the CMOS in terms of audio quality and speech prosody. We tell the tester to focus on one corresponding aspect and ignore the other aspect when scoring. We put more information about the subjective evaluation in Appendix~\ref{details_subjective_evaluation}.

\begin{table}[!t]
\caption{The objective and subjective comparisons for zero-shot text-to-speech synthesis. We evaluate the audio quality, speech prosody, and speaker similarity of different systems on the VCTK and LibriSpeech test-clean datasets with 95\% confidence intervals.}
\label{table_zero_shot}
\small
\centering
\begin{tabular}{@{}l|l|ccc|cc@{}}
\toprule
\multirow{2}{*}{Dataset} & \multirow{2}{*}{Method} & \multicolumn{3}{c|}{Subjective}           & \multicolumn{2}{c}{Objective}                \\ 
 &  & MOS-Q ($\uparrow$) & MOS-P ($\uparrow$) & MOS-S ($\uparrow$) & Pitch ($\downarrow$) & Speaker ($\uparrow$) \\ \midrule
\multirow{3}{*}{VCTK} & Ground Truth & 4.35 $\pm$ 0.11 & 4.48 $\pm$ 0.10 & 4.33 $\pm$ 0.13 & - & 0.915 \\
 & YourTTS~\citep{casanova2022yourtts} & 4.04 $\pm$ 0.10 & 4.18 $\pm$ 0.09 & 3.76 $\pm$ 0.12 & 32.43 & 0.847 \\
 & Mega-TTS   & \bfseries 4.27 $\pm$ 0.09 & \bfseries 4.32 $\pm$ 0.11 & \bfseries 4.27 $\pm$ 0.10 & \bfseries 17.45 & \bfseries 0.877 \\ \midrule
\multirow{3}{*}{LibriSpeech} & Ground Truth & 4.23 $\pm$ 0.13 & 4.49 $\pm$ 0.11 &  4.29$\pm$0.16 & - & 0.956 \\
 & YourTTS~\citep{casanova2022yourtts} & 3.83 $\pm$ 0.12 & 4.06 $\pm$ 0.13 & 3.22 $\pm$ 0.21 & 44.05 & 0.909                           \\
 & Mega-TTS & \bfseries 4.08 $\pm$ 0.17& \bfseries 4.21$\pm$0.17& \bfseries 3.90 $\pm$ 0.18      & \bfseries 35.46 & \bfseries 0.936 \\ \bottomrule
\end{tabular}
\end{table}

\begin{table}[!t]
\caption{The comparison between Mega-TTS and VALL-E.}
\label{table_vs_valle}
\small
\centering
\begin{tabular}{@{}l|ccc@{}}
\toprule
Method & CMOS-Q & CMOS-P & MOS-S ($\uparrow$) \\ \midrule

VALL-E~\citep{wang2023neural} & -0.23 & -0.27 & 4.06 $\pm$ 0.22 \\
Mega-TTS   & \bfseries 0.00 & \bfseries 0.00 & \bfseries 4.11 $\pm$ 0.21 \\ \bottomrule
\end{tabular}
\end{table}

\subsection{Results of zero-shot synthesis}
We compare the zero-shot synthesis performance of Mega-TTS with baseline systems, including: 1)YourTTS~\citep{casanova2022yourtts}, a powerful zero-shot TTS model trained on 1k hours of speech dataset. We use the official code and released checkpoint\footnote{\url{https://github.com/Edresson/YourTTS}}; 2) VALL-E, a large-scale zero-shot TTS model using the audio codec model to generate discrete speech codes and LLM to generate them. For VALL-E, we directly download the first 16 utterances from the VALL-E demo page. The audio samples consist of 8 samples from LibriSpeech and 8 samples from VCTK\footnote{VALL-E does not release its code officially. The unofficial implementations and our implementation are deficient, which would make us difficult to fairly compare our system with VALL-E.}. As shown in Table~\ref{table_zero_shot}, Mega-TTS significantly outperforms YourTTS in terms of audio quality and speech prosody. And in terms of speaker similarity, Mega-TTS significantly outperforms YourTTS with +0.51 MOS-S on VCTK and +0.68 MOS-S on LibriSpeech, demonstrating the effectiveness of Mega-TTS in zero-shot scenarios. Besides, as shown in Table~\ref{table_vs_valle}, Mega-TTS outperforms VALL-E in all metrics. It can be seen that Mega-TTS is able to generate more natural speeches than VALL-E, demonstrating the effectiveness of introducing intrinsic inductive biases. To further investigate the performance of disentanglement, we also visualize the distribution of the timbre and prosody representations in Appendix~\ref{app:visualization_representation}.

\begin{table}[!t]
\caption{The MOS evaluation ($\uparrow$) for speech quality, speech prosody, and speaker similarity on speech editing task on the VCTK dataset with 95\% confidence intervals.}
\centering
\small
\label{table_speech_editing}
\begin{tabular}{@{}l|ccc@{}}
\toprule
Method
& MOS-Q ($\uparrow$)
& MOS-P ($\uparrow$)
& MOS-S ($\uparrow$)\\ 
\midrule
EditSpeech~\citep{tan2021editspeech} & 3.57 $\pm$ 0.12  & 3.87 $\pm$ 0.14 & 3.93 $\pm$ 0.14 \\
A${}^3$T~\citep{bai20223} & 3.73 $\pm$ 0.13 & 3.96 $\pm$ 0.14 & 3.97 $\pm$ 0.12 \\
Mega-TTS & \bfseries 3.81 $\pm$ 0.14 & \bfseries 4.11 $\pm$ 0.14 & \bfseries 4.36 $\pm$ 0.16 \\ 
\bottomrule
\end{tabular}
\end{table}

\subsection{Results of zero-shot speech editing}
We compare the quality of generated audio samples of our Mega-TTS with SOTA speech editing baselines, including 1) EditSpeech~\citep{tan2021editspeech}; 2) A${}^3$T~\citep{bai20223}. Since the text content of the generated speech has been edited in the speech editing evaluation, the ground truth is missing. Therefore, we only conduct the subjective evaluation. We manually define modification operations (i.e., insertion, replacement, and deletion) of the test samples. We then conduct the experiments on the VCTK dataset. We evaluate the audio quality, speech prosody, and speaker similarity for each audio sample. The results are presented Table~\ref{table_speech_editing}. It can be seen that Mega-TTS achieves the highest perceptual quality, prosody, and speaker similarity score, which demonstrates the effectiveness of our proposed speech prompting mechanism for speech editing and the powerful in-context learning capability of Mega-TTS.

\begin{table}[!t]
\caption{The comparisons for cross-lingual text-to-speech synthesis with 95\% confidence intervals. We also measure the word error rate (WER) and speaker similarity score for the objective evaluations.}
\label{table_cross_lingual_tts}
\centering
\small
\begin{tabular}{@{}l|ccc|cc@{}}
\toprule
\multirow{2.5}{*}{ Method} & \multicolumn{3}{c|}{Subjective} & \multicolumn{2}{c}{Objective} \\ \cmidrule(l){2-6}  & MOS-Q ($\uparrow$)& MOS-P ($\uparrow$) 
                             & MOS-S ($\uparrow$) & WER ($\downarrow$) & Speaker ($\uparrow$) \\ \midrule
YourTTS~\citep{casanova2022yourtts} & 3.65 $\pm$ 0.21 & 3.92 $\pm$ 0.18 & 3.32 $\pm$ 0.27 & 7.59\% & 0.883 \\
VALL-E X~\citep{zhang2023speak} & 3.73 $\pm$ 0.17 & 3.97 $\pm$ 0.18 & 3.81 $\pm$ 0.16 & - & - \\
Mega-TTS & \bfseries 3.85 $\pm$ 0.17 & \bfseries 4.08 $\pm$ 0.19 & \bfseries 3.86 $\pm$ 0.18 & \bfseries 3.04\% & \bfseries 0.919 \\

\bottomrule
\end{tabular}
\end{table}

\subsection{Results of zero-shot cross-lingual TTS}
To compare Mega-TTS with the zero-shot cross-lingual TTS models VALL-E X~\citep{zhang2023speak}, we directly download the utterances from the VALL-E X demo page, which consists of 6 speech pairs from LibriSpeech, EMIME, and AISHELL-3. Since YourTTS~\citep{casanova2022yourtts} is built only for English TTS, we evaluate the performance of English TTS with Chinese samples as prompts. The results are listed in Table~\ref{table_cross_lingual_tts}. It can be seen that Mega-TTS surpasses VALL-E X in terms of audio quality, speech prosody, and speaker similarity scores, which further demonstrates the superiority of introducing proper inductive biases to different speech attributes. For objective evaluations, we use all of the text samples in the LibriSpeech test-clean set as the target sentences and randomly select one audio from AISHELL-3 as the speech prompt for each target sentence. The results show that Mega-TTS achieves a significantly lower WER than YourTTS, demonstrating the effectiveness of our method.

\begin{table}[!t]
\caption{The comparison of robustness between Mega-TTS and other systems on the 50 particularly hard sentences. Each kind of word error is counted at once per sentence.}
\label{table_robustness}
\centering
\small
\begin{tabular}{@{}l|c|c|c|c@{}}
    \toprule
    Method           & Repeats & Skips & Error Sentences & Error Rate        \\
    \midrule
    Tacotron~\citep{wang2017tacotron} & 10 & 16 & 22 & 44\% \\
    VALL-E~\citep{wang2023neural}    & 8 & 11 & 14 & 28\%\\
    FastSpeech~\citep{ren2019fastspeech} & 0 & 0 & 0 & 0\%\\

    Mega-TTS & 0 & 0 & 0 & 0\% \\ 
    \bottomrule
\end{tabular}
\end{table}

\subsection{Results of robustness evaluation}
To further evaluate the robustness of the proposed model, we adopt the 50 particularly hard sentences following FastSpeech~\citep{ren2019fastspeech}. As shown in Table~\ref{table_robustness}, Tacotron~\citep{wang2017tacotron} and VALL-E~\citep{wang2023neural} show poor robustness on these complicated sentences. As a comparison, our Mega-TTS shows equivalent robustness to the non-autoregressive models, such as FastSpeech~\citep{ren2019fastspeech}, without any repeat or skip issues. It can be seen that directly modeling the discrete speech tokens with LLMs like VALL-E~\citep{wang2023neural} would cause robustness issues. As a comparison, Mega-TTS not only leverages the in-context learning capability of LLMs, but also maintains good robustness by introducing the proper inductive bias to each speech component.

\section{Conclusion}
\label{conclusion}
In this paper, we proposed Mega-TTS, which aims to introduce proper inductive biases into large-scale zero-shot TTS systems. We disentangle speech into different attributes (i.e., content, timbre, prosody, and phase) and model different attributes in different ways. We train Mega-TTS with 20K hours of multi-domain speech data and evaluate its performance on unseen datasets. Our experimental results on three speech synthesis tasks show that Mega-TTS outperforms state-of-the-art zero-shot TTS models regarding audio quality, speech prosody, speaker similarity, and robustness. Due to limited page space, we discuss the limitations and future works in Appendix~\ref{app:limitation_and_future_work} and the broader impacts in Appendix~\ref{app:broader_impacts}.


\bibliographystyle{plain}
\bibliography{ref}

\newpage
\appendix
\appendixpage

\section{Detailed Experimental Settings}
\subsection{Model Configurations}
\label{megatts_config}
We list the model hyper-parameters of Mega-TTS in Table~\ref{table_5}.
\begin{table}[h]
\caption{Hyperparameters of Mega-TTS models.}
\label{table_5}
\centering
\begin{tabular}{@{}llcc@{}}
\toprule
\multicolumn{2}{c}{Hyper-parameter} & Value  \\ \midrule
\multicolumn{1}{l|}{\multirow{5}{*}{Prosody Encoder}}           & \multicolumn{1}{l|}{Encoder Layers}                 & 5    \\
\multicolumn{1}{l|}{}                                            & \multicolumn{1}{l|}{Hidden Size}                    & 320  \\
\multicolumn{1}{l|}{}                                            & \multicolumn{1}{l|}{Conv1D Kernel}                  & 5    \\ \multicolumn{1}{l|}{}                                            & \multicolumn{1}{l|}{VQ Embedding Size}              & 2048    \\ \multicolumn{1}{l|}{}                                            & \multicolumn{1}{l|}{VQ Embedding Channel}           & 256    \\ \midrule
\multicolumn{1}{l|}{\multirow{5}{*}{Content Encoder}}                            & \multicolumn{1}{l|}{Phoneme Embedding Size}         & 320  \\ \multicolumn{1}{l|}{}                                            & \multicolumn{1}{l|}{Encoder Layers}                 & 4  \\
\multicolumn{1}{l|}{}                                          & \multicolumn{1}{l|}{Hidden Size}                    & 320  \\
\multicolumn{1}{l|}{}                                          & \multicolumn{1}{l|}{Kernel Size}                  & 5  \\
\multicolumn{1}{l|}{}                                          & \multicolumn{1}{l|}{Filter Size}             & 1280  \\\midrule
\multicolumn{1}{l|}{\multirow{3}{*}{Timbre Encoder}}        & \multicolumn{1}{l|}{Encoder Layers}                    & 5  \\
\multicolumn{1}{l|}{}                                            & \multicolumn{1}{l|}{Hidden Size}                    & 320  \\
\multicolumn{1}{l|}{}                                            & \multicolumn{1}{l|}{Conv1D Kernel}                  & 31   \\ \midrule
\multicolumn{1}{l|}{\multirow{3}{*}{Mel Decoder}}      &  \multicolumn{1}{l|}{Decoder Layers}                 & 5    \\
\multicolumn{1}{l|}{}                                            & \multicolumn{1}{l|}{Hidden Size}                    & 320    \\
\multicolumn{1}{l|}{}                                            & \multicolumn{1}{l|}{Conv1D Kernel}                  & 5  \\ \midrule
\multicolumn{1}{l|}{\multirow{7}{*}{P-LLM}}      & \multicolumn{1}{l|}{Decoder Layers}                                 & 8    \\
\multicolumn{1}{l|}{}                                            & \multicolumn{1}{l|}{Hidden Size}                 & 512    \\
\multicolumn{1}{l|}{}                                            & \multicolumn{1}{l|}{Decoder Kernel Size}         & 5  \\
\multicolumn{1}{l|}{}                                            & \multicolumn{1}{l|}{Decoder channel size}        & 2048   \\
\multicolumn{1}{l|}{}                                            & \multicolumn{1}{l|}{Prosody Code Embedding Size} & 2050    \\
\multicolumn{1}{l|}{}                                            & \multicolumn{1}{l|}{Attention Headss} & 8    \\
\multicolumn{1}{l|}{}                                            & \multicolumn{1}{l|}{Number of Contextual Sentences} & 7    \\\midrule
\multicolumn{1}{l|}{\multirow{4}{*}{Multi-Length Discriminator}} & \multicolumn{1}{l|}{Number of Discriminators}       & 3    \\
\multicolumn{1}{l|}{}                                            & \multicolumn{1}{l|}{Window Size}                    & 32, 64, 128 \\
\multicolumn{1}{l|}{}                                            & \multicolumn{1}{l|}{Conv2D Layers}                  & 3           \\
\multicolumn{1}{l|}{}                                            & \multicolumn{1}{l|}{Hidden Size}                    & 192         \\ \midrule
\multicolumn{2}{c|}{Total Number of Parameters}                                                                                                           & 222.5M                                  \\ \bottomrule
\end{tabular}
\end{table}

\begin{figure}[tbp]
    \centering
	\begin{minipage}{0.85\linewidth}
		\centering
		\includegraphics[width=1\linewidth]{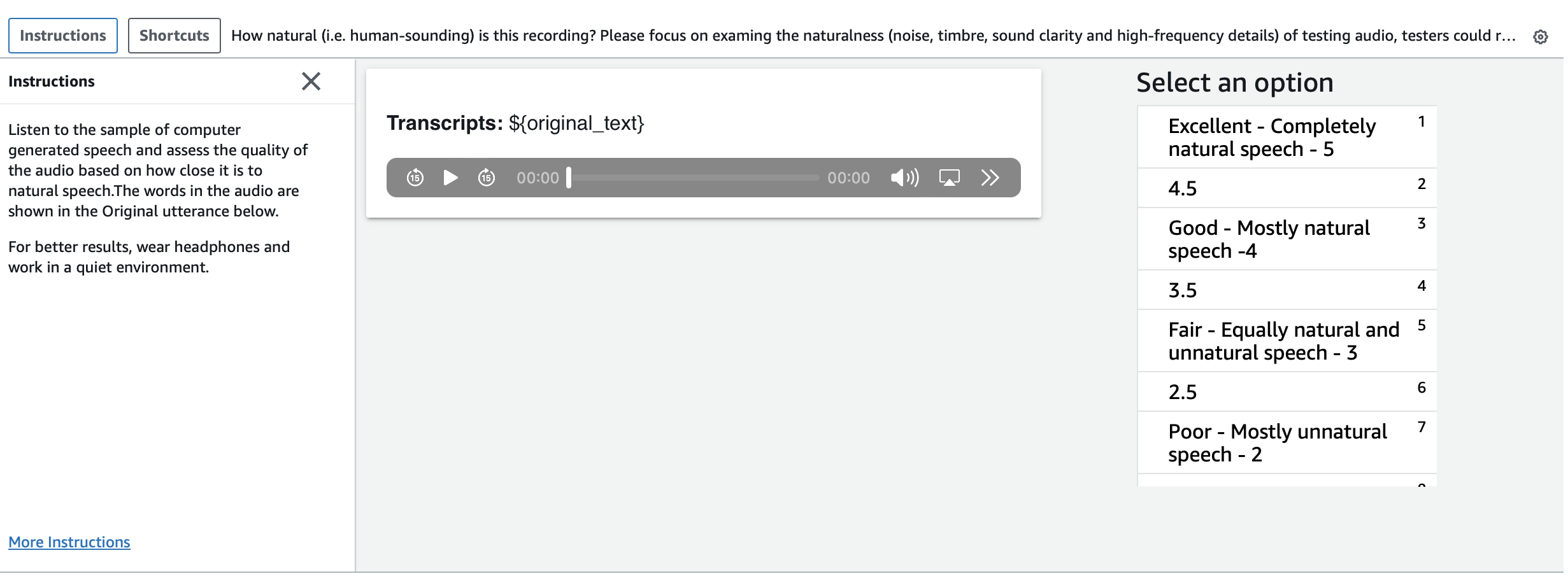}
		\caption*{(a)  Screenshot of MOS-Q testing.}
	\end{minipage}
	\centering
	\begin{minipage}{0.85\linewidth}
		\centering
		\includegraphics[width=1\linewidth]{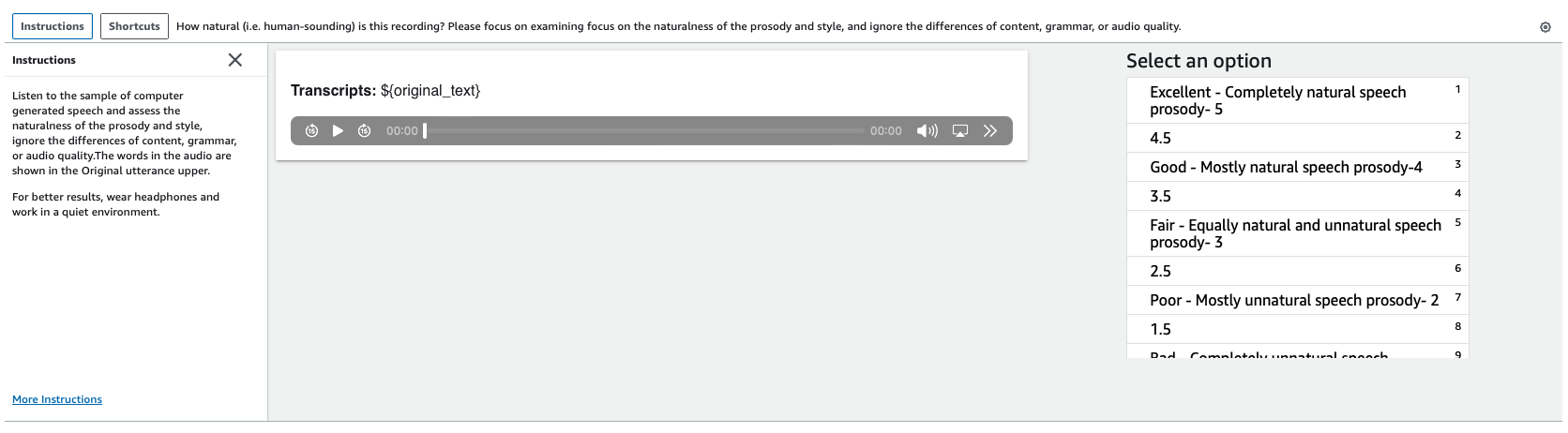}
		\caption*{(b) Screenshot of MOS-P testing.}
	\end{minipage}
	\centering
	\begin{minipage}{0.85\linewidth}
		\centering
		\includegraphics[width=1\linewidth]{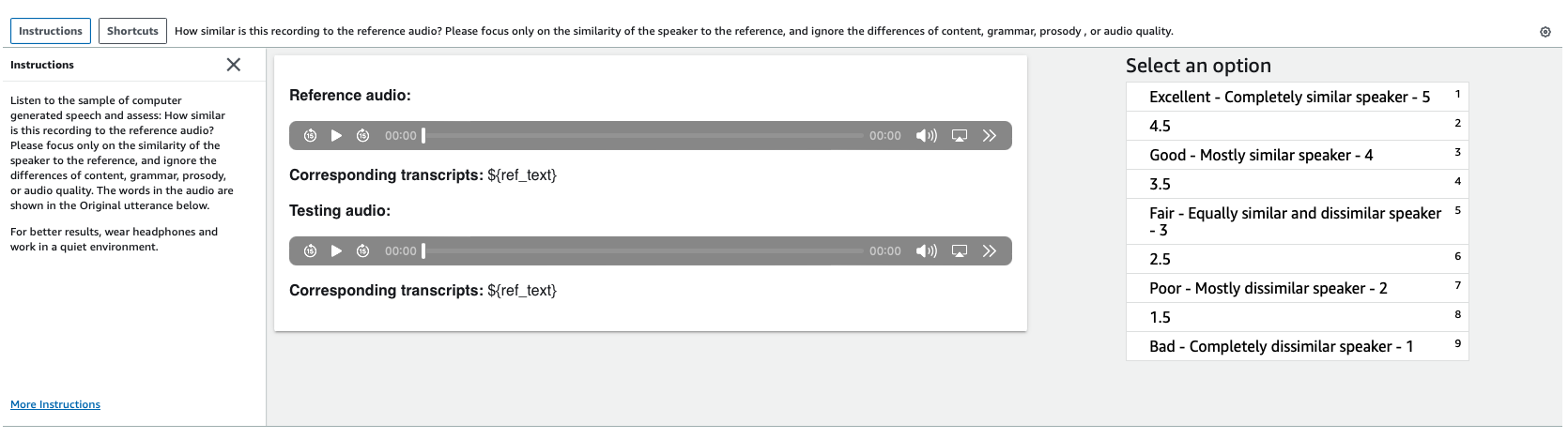}
		\caption*{(c) Screenshot of MOS-S testing.}
	\end{minipage}
	\centering
	\begin{minipage}{0.85\linewidth}
		\centering
		\includegraphics[width=1\linewidth]{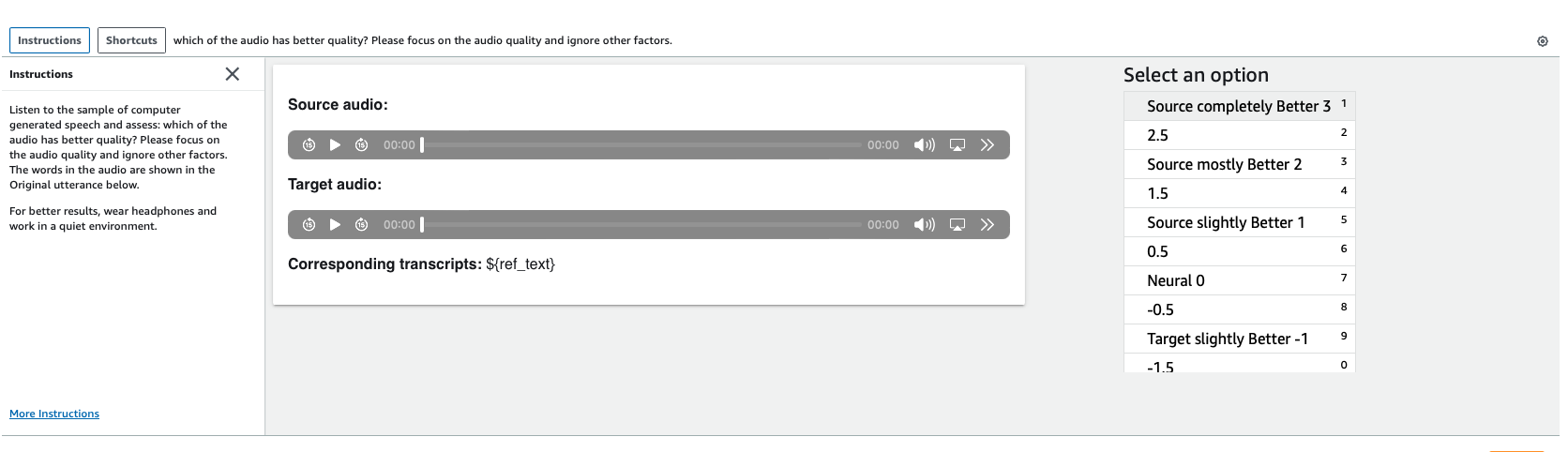}
		\caption*{(d) Screenshot of CMOS-Q testing.}
	\end{minipage}
	\centering
    \begin{minipage}{0.85\linewidth}
		\centering
		\includegraphics[width=1\linewidth]{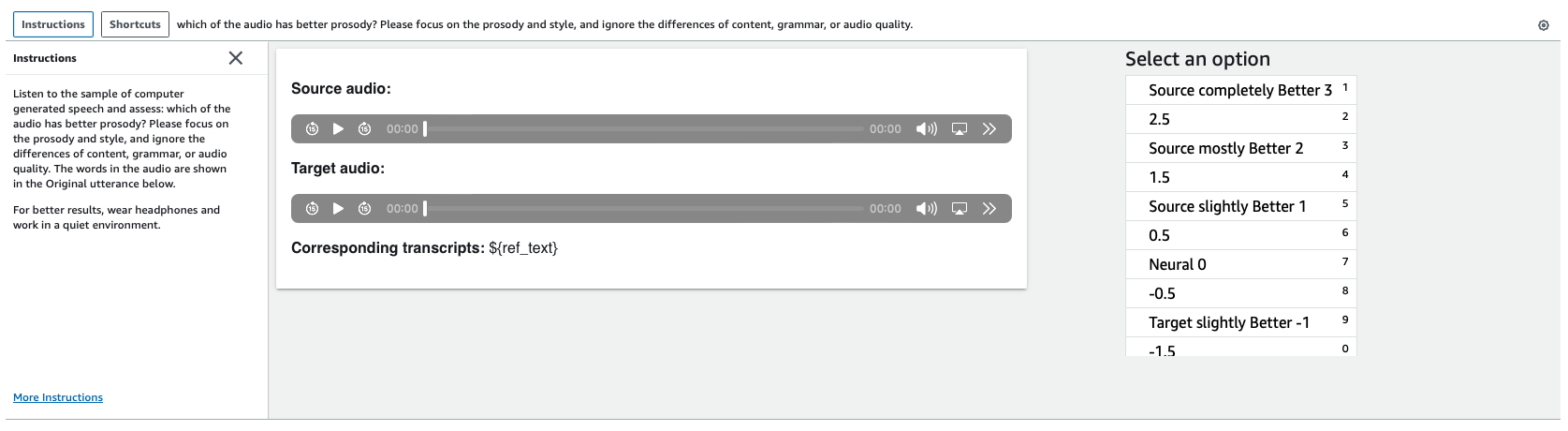}
		\caption*{(e) Screenshot of CMOS-P testing.}
	\end{minipage}
	\centering
	\caption{Screenshots of MOS and CMOS testings for audio quality, speech prosody, and speaker similarity evaluations.}
	\label{screenshots_audio_quality_prosody}
\end{figure}

\subsection{Details in Subjective Evaluation}
\label{details_subjective_evaluation}
We perform the audio quality, speech prosody, and speaker similarity evaluations on Amazon Mechanical Turk (MTurk). For each dataset, we randomly select 50 samples from the test set and use the TTS systems to generate the audio samples. Each audio has been listened to by at least 20 listeners. For MOS, each tester is asked to evaluate the subjective score of a sentence on a 1-5 Likert scale. For CMOS, listeners are asked to compare pairs of audio generated by systems A and B, indicating which of the two audio they prefer, and choose one of the following scores according to the degree of superiority: 0 indicating no difference, 1 indicating source slightly better, 2 indicating source mostly better and 3 indicating source completely better. For audio quality evaluation (MOS-Q and CMOS-Q), we tell listeners to ``\textit{Please focus on the audio quality and ignore other factors}''. For prosody evaluations (MOS-P and CMOS-P), we tell listeners to ``\textit{Please focus on the prosody and style, and ignore the differences of grammar, audio quality, or other factors. }''. For speaker similarity evaluations (MOS-S), we tell listeners to ``\textit{ Please focus only on the similarity of the speaker to the reference, and ignore the differences of content, grammar, prosody, audio quality, or other factors.}''.

The screenshots of instructions for testers are shown in Figure~\ref{screenshots_audio_quality_prosody}. We paid \$12 to participants hourly and totally spent about \$1000 on participant compensation. We tell the participants that the data will be used in scientific research.

\subsection{Details of Speaker Diarization Model}
\label{app:diarization_model}
To obtain the speaker information from GigaSpeech and WenetSpeech, we use a released automatic speaker diarization model called \textit{pyannote.audio}\footnote{\url{https://huggingface.co/pyannote/speaker-diarization}}, which achieves DER=11.24\% on the VoxConverse dataset and DER=14.09\% on the AISHELL-4 dataset. We only assign the speaker ID to the audio clip when its activation score is higher than 70\% and abandon other audio clips. We also abandon the audio clips that contain multiple speakers speaking simultaneously.

\subsection{Details of Speaker Similarity Model}
\label{app:details_speaker_similarity_model}
To measure the speaker similarity, we use the WavLM~\citep{chen2022wavlm} model finetuned for speaker verification from \url{https://huggingface.co/microsoft/wavlm-base-plus-sv} to extract the speaker embedding. Then the cosine similarity between the synthesized speech's speaker embedding and the ground-truth speech's speaker embedding is calculated as the speaker similarity score. The WavLM model is pretrained on 94,000 hours of speech data and finetuned on the VoxCeleb1 dataset using an X-Vector head with an Additive Margin Softmax loss, which achieves 0.84\%, 0.928\%, and 1.758\% EER (Equal Error Rate) on the Vox1-O, Vox1-E, and Vox1-H trial lists.

\subsection{Details of ASR Model}
\label{app:details_asr_model}
To measure the audio quality and speech intelligibility for cross-lingual TTS systems, we evaluate the word error rate (WER) metric. We use the finetuned HuBERT-Large model to transcribe the synthesized speech into text and calculate the WER between the transcribed text and the original target text. The finetuned HuBERT-Large model from \url{https://huggingface.co/facebook/hubert-large-ls960-ft} is finetuned on 960h of Librispeech and achieves 1.5\%, 3.0\%, 1.9\%, and 3.3\% WER on the dev-clean, dev-other, test-clean, and test-other set of Librispeech.

\subsection{Error Bars and Random Seeds}
For the subjective evaluations, we report confidence intervals of the results of MOS tests in Table~\ref{table_zero_shot}, Table~\ref{table_vs_valle}, Table~\ref{table_speech_editing}, and Table~\ref{table_cross_lingual_tts}. For the objective evaluations, we ran the experiments 10 times with 10 different random seeds ($[1234,1111,2222,3333,4444,5555,6666,7777,8888,9999]$) and obtained the averaged results.

\section{Visualizations of Mel-Spectrograms}
\begin{figure}[tbp]
	\centering
	\begin{minipage}{0.32\linewidth}
		\centering
		\includegraphics[width=1\linewidth]{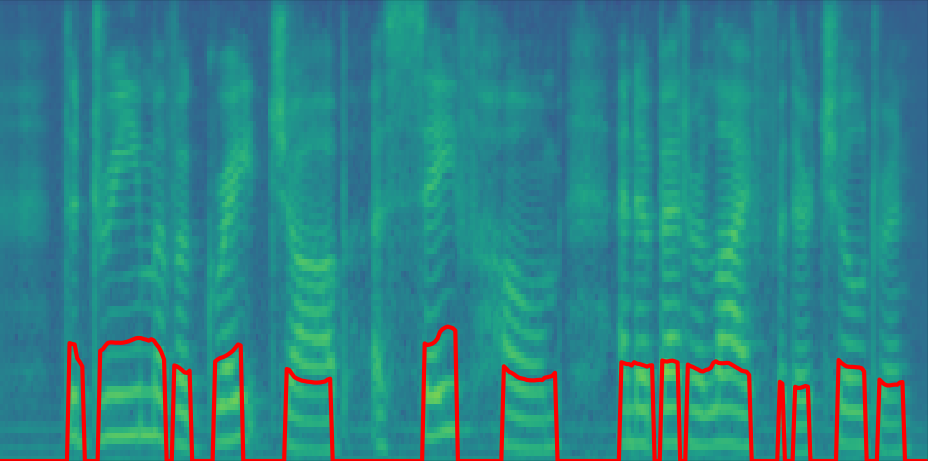}
		\caption*{(a) S=1234}
	\end{minipage}
	\centering
	\begin{minipage}{0.32\linewidth}
		\centering
		\includegraphics[width=1\linewidth]{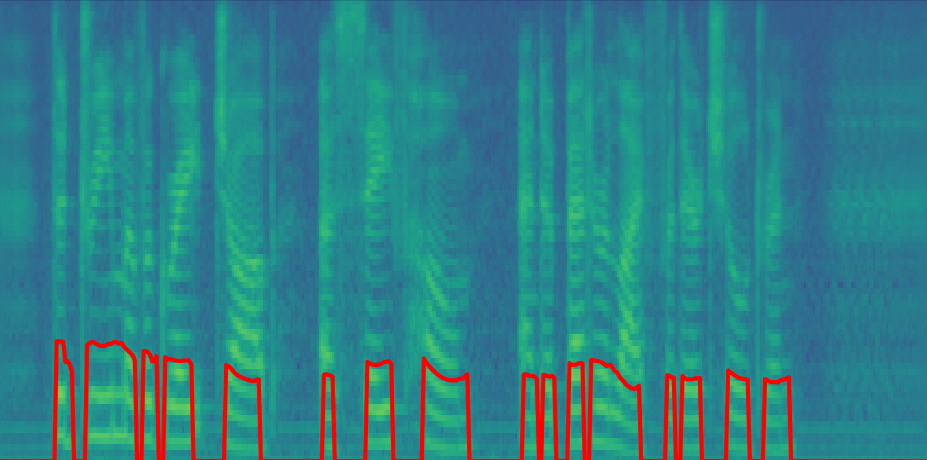}
		\caption*{(b) S=1334}
	\end{minipage}
	\centering
	\begin{minipage}{0.32\linewidth}
		\centering
		\includegraphics[width=1\linewidth]{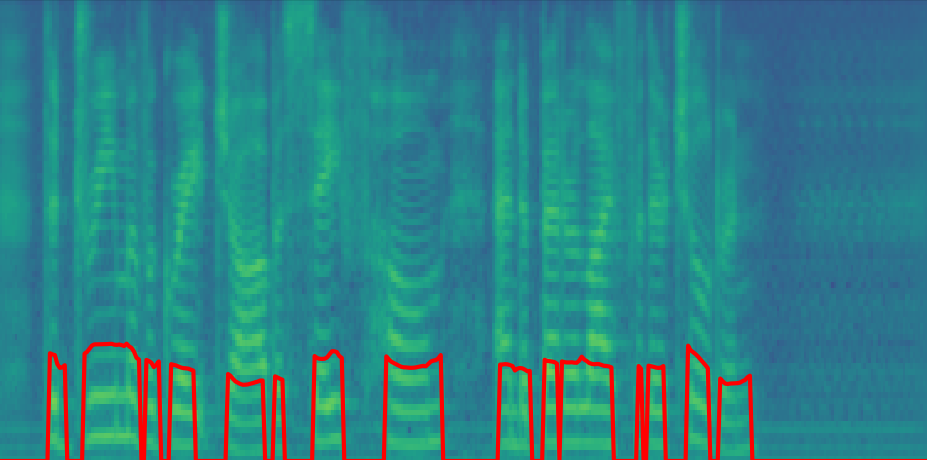}
		\caption*{(c) S=1734}
	\end{minipage}
	\centering
	\begin{minipage}{0.32\linewidth}
		\centering
		\includegraphics[width=1\linewidth]{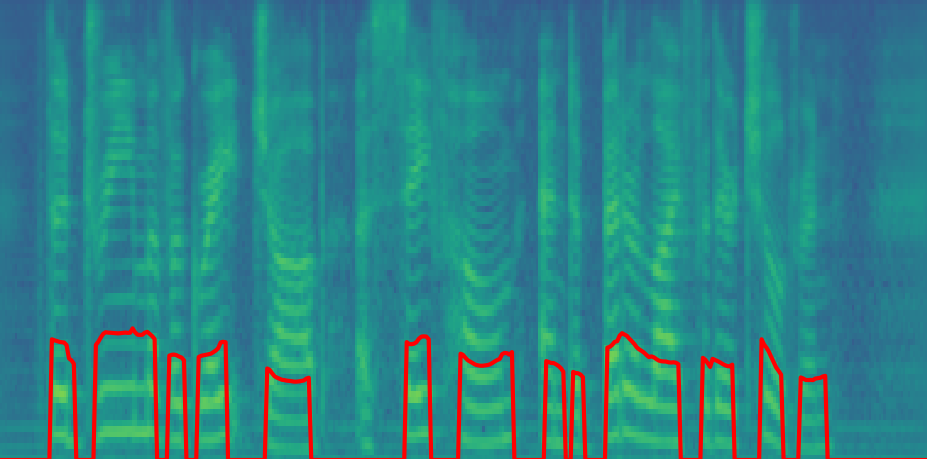}
		\caption*{(d) S=3234}
	\end{minipage}
	\centering
	\begin{minipage}{0.32\linewidth}
		\centering
		\includegraphics[width=1\linewidth]{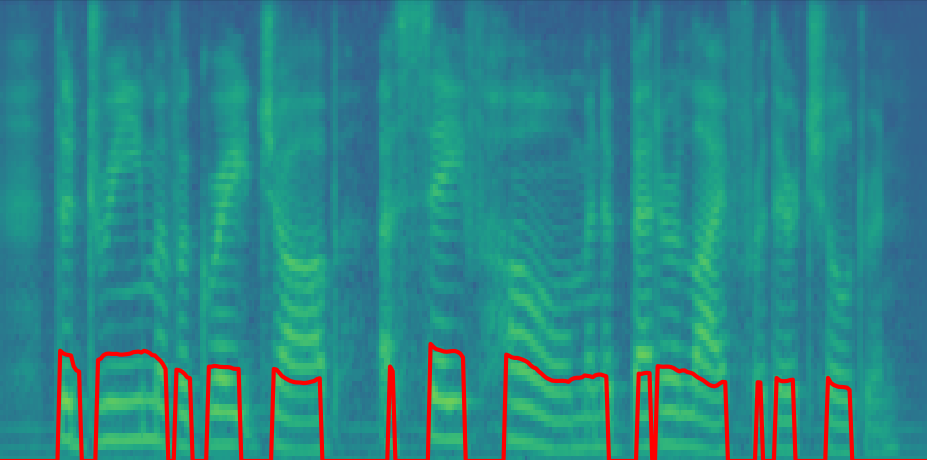}
		\caption*{(e) S=4999}
	\end{minipage}
	\centering
	\begin{minipage}{0.32\linewidth}
		\centering
		\includegraphics[width=1\linewidth]{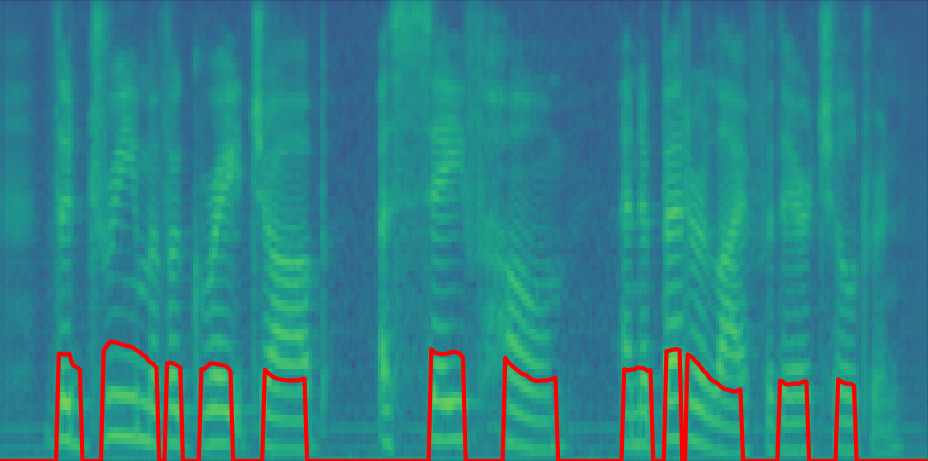}
		\caption*{(f) S=9222}
	\end{minipage}
	\centering
	\caption{Visualizations of the mel-spectrograms generated with different random seeds S.}
	\label{vis_spec}
\end{figure}

We put more visualizations of mel-spectrograms with different random seeds in Figure~\ref{vis_spec}. We can see that with different random seeds, Mega-TTS can generate diverse results that have different prosody and frequency details.
\label{app:vis_mel}

\section{Visualization of Representations}
\label{app:visualization_representation}
\begin{figure*}[ht]
	\centering
	\includegraphics[width=0.65\textwidth]{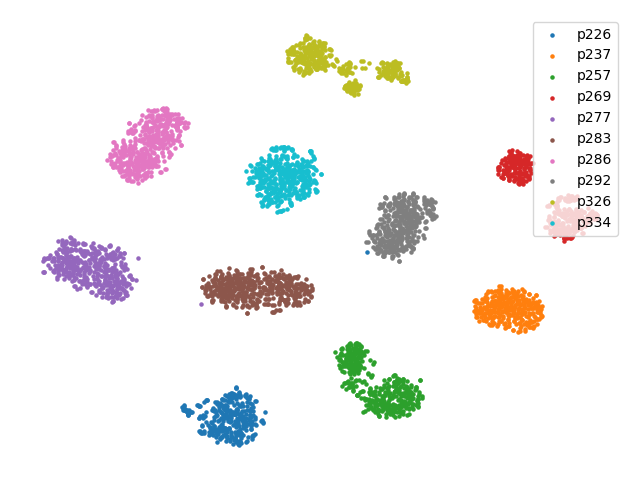}
	\caption{The T-SNE visualization of timbre embeddings for 10 unseen speakers on the VCTK dataset.  }
	\label{fig:tsne_1}
\end{figure*}

\begin{figure*}[ht]
	\centering
	\includegraphics[width=0.65\textwidth]{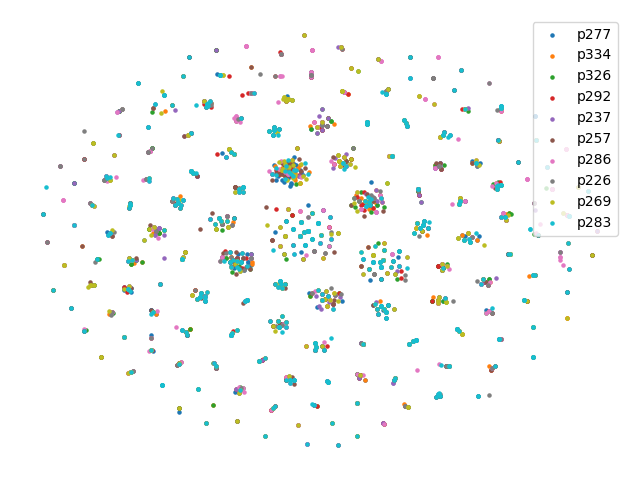}
	\caption{The T-SNE visualization of prosody embeddings for 10 unseen speakers on the VCTK dataset.  }
	\label{fig:tsne_2}
\end{figure*}

To validate the effectiveness of disentanglement for speech components in Section~\ref{sec_3_1}, we adopt T-SNE~\citep{van2008visualizing} to visualize timbre embedding and prosody embedding for unseen speakers on the VCTK dataset. We randomly select 10 speakers and directly use the encoders proposed in Section~\ref{sec_3_1} to extract the timbre and prosody information from their audio samples. The results are shown in Figure~\ref{fig:tsne_1} and Figure~\ref{fig:tsne_2}. It can be seen that the timbre embeddings are ideally located according to the speaker ID. However, the prosody embeddings of different speakers have similar distributions. It shows that our proposed prosody and timbre encoders are able to disentangle the corresponding representations from the mel-spectrograms, which further ensures the effectiveness of our P-LLM.


\begin{table}[!t]
\caption{The disentanglement performance evaluation for VQGAN-based TTS model with different VQ hyperparameters.}
\label{table:hyperparameter_selection_for_bn}
\centering
\begin{tabular}{@{}c|cc@{}}
\toprule
Channel Size * Embedding Size & Pitch ($\downarrow$) & Speaker ($\uparrow$)\\ \midrule
64*512     & 73.82 & 0.719 \\
256*2048  & \bfseries 49.30 & \bfseries0.941   \\
1024*4096 & 78.84 & 0.707 \\
\bottomrule
\end{tabular}
\end{table}

\section{Hyperparameter Selection for the Information Bottleneck}
\label{app:hyperparameter_selection_for_bn}
In this section, we describe the details of the hyperparameter selection for the information bottleneck proposed in Section~\ref{sec_3_1}. The information bottleneck of Mega-TTS mainly contains two key hyperparameters: the channel size and the embedding size of the vector quantization (VQ) layer. When the channel size and the embedding size are too small or large, the performance of disentanglement will be poor. Therefore, we should carefully select these hyperparameters. We train the VQGAN-based TTS models with different VQ hyperparameters and evaluate their pitch distance and speaker similarity following Section~\ref{sec_4}. Differently, we use the proposed encoders to extract the timbre, content, and prosody embeddings of the test samples. Then, we randomly shuffle the timbre embedding sequence and reconstruct the mel-spectrogram with the original content, original prosody, and shuffled timbre information. We calculate the pitch distance between the ground-truth speech and the generated speech, but we calculate the speaker similarity between the shuffled ground-truth speech and the generated speech. As shown in Table~\ref{table:hyperparameter_selection_for_bn}, when the channel size is 256 and the embedding size is 2048, the VQGAN-based TTS model shows the best pitch accuracy and speaker similarity, i.e., the disentanglement performance is the best.

\section{Ablation Studies of Dataset Size and Model Size}
\label{app:ablation_studies_of_dataset_size_and_model_size}
In this section, we evaluate the influences of the training dataset size and model size on the zero-shot TTS task for Mega-TTS. We evaluate the pitch distance, speaker similarity, and the average absolute duration error in milliseconds on the LibriSpeech test-clean set. As shown in Table~\ref{table:different_size_training_data}, when the dataset size grows, the zero-shot performance of Mega-TTS is significantly improved. Moreover, from Table~\ref{table:different_hidden_size_pllm}, we can see that when the hidden size of P-LLM grows, the pitch distance significantly drops, demonstrating that the in-context learning capability of P-LLM can be greatly improved by the size of the model.

\begin{table}[!t]
\caption{The performance of Mega-TTS using different sizes of the training data.}
\label{table:different_size_training_data}
\centering
\begin{tabular}{@{}c|c|ccc@{}}
\toprule
Dataset Usage & Total Time (hours) & Pitch ($\downarrow$) & Speaker ($\uparrow$)  & Duration ($\downarrow$) \\ \midrule
GiGaSpeech    & 10K  & \textbf{36.50} & \textbf{0.935} & \textbf{62.61} \\
LibriSpeech   & 960  & 43.90 & 0.915 & 69.85  \\
VCTK          & 44   & 81.33 & 0.828 & 82.39 \\
\bottomrule
\end{tabular}
\end{table}

\begin{table}[!t]
\caption{The performance of Mega-TTS with different hidden sizes of P-LLM.}
\label{table:different_hidden_size_pllm}
\centering
\begin{tabular}{@{}c|cc@{}}
\toprule
Hidden Size of P-LLM & Pitch ($\downarrow$) & Speaker ($\uparrow$) \\ \midrule
128   & 82.24 & 0.917  \\
256   & 71.74 & 0.920 \\
512   & \textbf{35.46} & \textbf{0.936} \\
\bottomrule
\end{tabular}
\end{table}





\section{Limitations and Future Works}
\label{app:limitation_and_future_work}
Although achieving superior performance on various zero-shot speech synthesis tasks, Mega-TTS still suffers from two main limitations.
\paragraph{Data coverage.} Although we use 20K hours of multi-domain data for training, our model still cannot cover everyone's voice. Especially for some speakers with extremely heavy accents, our model cannot imitate their speaking style very well. In the future, we will further scale up the training data to 200K hours to further improve the performance of the model.
\paragraph{Reconstruction Robustness.} Although the reconstruction quality of the proposed VQGAN-based TTS model is satisfying on the clean dataset, it will be influenced by the background music or the extremely loud reverberation. In future work, we will explore a new model structure that is more robust against the acoustic environment noises.

\section{Broader Impacts} 
\label{app:broader_impacts}
Mega-TTS improves the quality and efficiency of zero-shot speech synthesis, which makes it easier for people to synthesize personalized speeches. In most cases, people will utilize this technique to facilitate movies, games, podcasts, and other services only. However, it may carry potential risks in misuse of the model, such as spoofing voice or other deepfake-related usages. To handle this, potential solutions like building a corresponding deepfake detection model should be considered. We also plan to include restrictions in the open-source license of the Mega-TTS
project to prevent the misuse of the model.

\end{document}